\documentclass[preprint,preprintnumbers,amssymb,superscriptaddress,aps,prd,nofootinbib,11pt]{revtex4-1}

\selectfont 

\usepackage{graphicx}       
\usepackage{dcolumn}        
\usepackage{bm}             
\usepackage{psfrag}
\usepackage{tensor}
\usepackage[usenames]{color}
\definecolor{navyblue}{rgb}{0.0, 0.0, 0.5}
\usepackage[linktocpage,colorlinks=true,allcolors=navyblue]{hyperref}
\usepackage{amsmath}
\usepackage{subfigure}

\newcommand{\beq}{\begin{equation}}
\newcommand{\eeq}{\end{equation}}

\begin{document}

\preprint{}

 \title{Comment on ``Geodesic dynamics on Chazy-Curzon spacetimes''}

\author{Sam R. Dolan}
 \email{s.dolan@sheffield.ac.uk}
\affiliation{Consortium for Fundamental Physics,
School of Mathematics and Statistics,
University of Sheffield, Hicks Building, Hounsfield Road, Sheffield S3 7RH, United Kingdom}

\date{\today}

\begin{abstract}
The recent numerical results of Dubeibe \emph{et al.} [arXiv:1812.08663] were interpreted as hinting at the existence of a fourth constant of motion (``Carter's constant'') for geodesics on Chazy-Curzon spacetimes. Here we show that, to the contrary, the geodesic dynamics of the single-particle Chazy-Curzon spacetime exhibit two features of a non-integrable system: chaotic orbits in the meridian plane, and Birkhoff chains in the surface of section. Thus, one should not expect Liouville-integrability, nor a fourth constant, for this system.
\end{abstract}

\pacs{}
\maketitle

%
%

\section{Introduction}
In Ref.~\cite{Dubeibe:2018tnv}, Dubeibe \emph{et al.}~presented Poincar\'e sections for timelike geodesic orbits on Chazy-Curzon (CC) spacetimes \cite{Chazy:1924,Curzon:1925,Griffiths:2009exact} of both one-particle and two-particle types. Observing regular sections, the authors conjectured that ``our findings suggest the existence of the so-called Carter's constant in both systems, which shall provide the fourth conserved quantity necessary to uniquely determine all orbits''. The purpose of this note is to show that, to the contrary, geodesic dynamics of the single-particle Chazy-Curzon spacetime bears the expected hallmarks of a non-integrable system (and see \cite{Sota:1995ms} for consideration of the two-particle case). Thus, one should not anticipate the existence of an additional (fourth) constant of motion polynomial in the momenta. In other words, geodesic motion on the CC spacetime is not Liouville integrable.

The question of the integrability of geodesic equations has been considered by several others in the wider context of Stationary Axisymmetric Vacuum (SAV) spacetimes \cite{Sota:1995ms,Gair:2007kr, Brink:2009mq,Brink:2009mt,Kruglikov:2011ky,Maciejewski:2013ncb,Vollmer:2016qpd}. The twin properties of stationarity and axisymmetry imply that every SAV spacetime admits a pair of Killing vector fields, and thus a pair of constants of motion corresponding to energy $E$ and azimuthal angular momentum $L$, respectively. Thus, the study of geodesic motion reduces to the study of a 2D Hamiltonian $H(\rho,z , p_\rho, p_z; E, L)$ in the meridian plane $(\rho, z)$. The existence of a fourth independent constant of motion (beyond $E$, $L$ and $H$ itself) would bring about Liouvillian integrability with respect to the canonical Poisson bracket. Remarkably, the Kerr spacetime -- a SAV spacetime describing a rotating black hole -- admits a rank-two irreducible Killing tensor, and thus a fourth constant of motion (`Carter's constant' \cite{Carter:1968rr}) that is quadratic in the momenta. A natural question is whether Carter-type constants exist in other SAV settings.

Geodesic motion on Zipoy-Voorhees (ZV) spacetimes \cite{Zipoy:1966,Voorhees:1971wh,Griffiths:2009exact}, also known as $\gamma$-metrics, has been studied from both numerical \cite{Sota:1995ms,Brink:2008xy,LukesGerakopoulos:2012pq,Lukes-Gerakopoulos:2013iia} and analytical \cite{Brink:2009mq,Brink:2009mt,Kruglikov:2011ky,Maciejewski:2013ncb,Vollmer:2016qpd} perspectives. The ZV metrics, with a parameter $\delta$, include as special cases both Minkowksi flat spacetime ($\delta = 0$) and Schwarzschild spacetime ($\delta = 1$). The single-particle CC spacetime emerges from the limit $\delta \rightarrow \infty$. A key lesson from investigations of ZV metrics is that finding apparent regularity in geodesics across large parts of the parameter space (and approximate constants of motion) is not, in itself, compelling evidence for Liouville integrability, that is, for the existence of a fourth constant.

In 1996, Sota \emph{et al.}~\cite{Sota:1995ms} showed, by numerical means, that the CC and ZV spacetimes exhibit apparently-regular dynamics in swathes of parameter space, for generic values of $\delta$. After further investigation \cite{Brink:2008xy}, Brink conjectured that the ZV metrics may give rise to integrable systems for values of $\delta$ other than $0$ and $1$. However, attempts to find the missing integral of geodesic motion led in the opposite direction. Brink showed \cite{Brink:2009mq,Brink:2009mt} that all SAV metrics admitting irreducible second-order Killing tensors are necessarily of Petrov type D. This key result implies that generic ZV metrics, which are not Petrov type D in general, do not allow for any new constant of geodesic motion which is quadratic in momentum variables. Exhaustive searches for higher-order irreducible Killing tensors on ZV spacetimes did not find any \cite{Brink:2009mt,Kruglikov:2011ky,Vollmer:2015nva,Vollmer:2016qpd}. 

In 2012, a numerical investigation by Lukes-Gerakopoulos \cite{LukesGerakopoulos:2012pq} showed that the Hamiltonian phase space of the ZV system has the features expected of a perturbed non-integrable system, such as chaotic layers, Birkhoff chains, and `stickiness' in the rotation number \cite{LukesGerakopoulos:2012pq,Lukes-Gerakopoulos:2013iia}. Chaotic layers are most visible near to the Lyapunov orbit, i.e., near the unstable periodic orbit at the threshold between bound orbits and plunging orbits. It is here that we shall focus our search for the signs of non-integrability in the CC case.

\section{Analysis and results}

\subsection{Geodesic equations}
The Chazy-Curzon spacetimes are part of the class of Static Axisymmetric Vacuum (StAV) metrics, itself a subclass of SAV metrics (see e.g.~Chap.~10 in Ref.~\cite{Griffiths:2009exact}). In cylindrical polar coordinates $\{t,\rho,z,\phi\}$, the StAV line element is typically expressed in Weyl form as
\beq
ds^2 = g_{\mu \nu} dx^\mu dx^\nu = -e^{2 \psi} dt^2 + e^{-2 \psi} \left( e^{2 \gamma} (d\rho^2 + dz^2) + \rho^2 d\phi^2 \right). \label{eq:weyl}
\eeq
Geodesic motion is governed by a Hamiltonian $H_4(x^\mu, p_\nu) = \frac{1}{2} g^{\mu \nu} p_\mu p_\nu = -\frac{1}{2}$ with $g^{\mu \nu}$ the metric inverse. $H_4$ is a constant of motion, which we set to $-\frac{1}{2}$ for test bodies of unit mass. The momentum variables are given by $p_\mu \equiv g_{\mu\nu} \dot{x}^\nu$, where the overdot denotes differentiation with respect to proper time $\tau$, i.e.,~$\dot{x}^\nu \equiv \frac{dx^\nu}{d \tau}$. 

It follows immediately from Hamilton's equations that $\dot{p}_{t} = 0 = \dot{p}_{\phi}$, as the metric inverse is not a function of $t$ or $\phi$, and thus we obtain two constants of motion, $E \equiv -p_t$ and $L = p_\phi$. Motion in the meridian plane is governed by a reduced Hamiltonian $H_2$, with the restriction $H_2 = 0$, given by
\beq
H_2 =  \Omega^2(\rho, z) \left( \frac{1}{2} \left( p_{\rho}^2 + p_{z}^2 \right) + V_{\text{eff}}(\rho,z; E, L) \right) , 
\eeq
where $\Omega \equiv \exp(\psi - \gamma)$, $V_{\text{eff}} \equiv -\frac{1}{2} \Omega^{-4} \Phi$ with $\Phi \equiv e^{-2\gamma} \left( E^2 - e^{2 \psi} - \rho^{-2} e^{4 \psi} L^2 \right)$  the effective potential defined in Eq.~(16) of Ref.~\cite{Dubeibe:2018tnv}. The contour $V_{\text{eff}} = 0$ defines a curve of zero velocity \cite{LukesGerakopoulos:2012pq}. The geodesics are found by solving Hamilton's equations, 
\beq
\dot{\rho} = \frac{\partial H_2}{\partial p_\rho}, \quad \dot{z} = \frac{\partial H_2}{\partial p_z}, \quad \dot{p}_\rho = - \frac{\partial H_2}{\partial \rho}, \quad \dot{p}_z = - \frac{\partial H_2}{\partial z}, \label{eq:hamiltons}
\eeq
with a choice of parameters $E$ and $L$ and initial conditions for $\rho, z, p_\rho, p_z$ such that the constraint $H_2 = 0$ is satisfied.

\subsection{Chazy-Curzon spacetime}
Einstein's vacuum field equations for line element (\ref{eq:weyl}) reduce to three equations: (i) Laplace's equation $\nabla^2 \psi = 0$, with $\nabla^2 \equiv \partial_{\rho \rho}+ \rho^{-1}\partial_\rho + \partial_{zz}$ the flat-space Laplacian; (ii) $\gamma_{,\rho} =  \rho (\psi_{,\rho}^2 - \psi_{,z}^2)$; and (iii) $\gamma_{,z} = 2 \rho \psi_{,\rho} \psi_{,z}$.

An elementary solution of Laplace's equation, for the potential of a `point source', yields
\beq
\psi = - \frac{M}{r} , \quad \quad \gamma = - \frac{M^2 \rho^2}{2 r^4}  , \label{eq:psigamma}
\eeq
where $r = \sqrt{\rho^2 + z^2}$ and $M$ is a constant. This is the single-particle Chazy-Curzon solution \cite{Chazy:1924,Curzon:1925,Griffiths:2009exact}. The metric (\ref{eq:weyl}) with functions (\ref{eq:psigamma}) is \emph{not} spherically symmetric and the interpretation of the causal and singularity structure of the resulting spacetime is somewhat involved \cite{Scott:1985ad,Scott:1985cb}. 

\subsection{Signatures of non-integrability}
At moderately large radii ($r \gtrsim 4M$), the CC system may be regarded as a perturbed (or `bumpy' \cite{Collins:2004ex}) Schwarzschild spacetime. The Schwarzschild spacetime is spherically symmetric, and thus it admits an additional Killing vector yielding a fourth constant of motion. The CC system resembles a Schwarzschild spacetime endowed with even mass multipoles ($\ell = 2n$, $n \in \mathbb{N}$), which break the spherical symmetry, and whose influence on the gravitational potential falls off as $r^{-(2 n +1)}$. In the absence of deeper symmetries, one would expect orbits on the CC system to exhibit the classic signatures of perturbed non-integrability.

In an \emph{integrable} Hamiltonian system with two degrees of freedom ($\rho$ and $z$), motion is restricted to a two-dimensional \emph{invariant torus} in the 4D phase space $(\rho,z,p_\rho,p_z)$. This motion has a pair of fundamental frequencies. If the ratio of frequencies is a rational number, then the torus is called resonant, and it has infinitely many periodic orbits with the same frequency ratio. Conversely, if the frequency ratio is irrational, a single orbit covers the whole of the non-resonant torus. When an integrable system is perturbed, the transition from integrability to non-integrability obeys two theorems: the Kolmogorov-Arnold-Moser (KAM) theorem and the Poincar\'e-Birkhoff (PB) theorem. According to the KAM theorem, for small perturbations, most non-resonant invariant tori of the integrable system are deformed yet survive in the perturbed system. These deformed tori are called KAM tori. The situation is different for the periodic orbits on the resonant tori, however. According to the PB theorem, out of the infinitely many periodic orbits on a resonant torus of the integrable system, only a finite even number survive in the perturbed system \cite{LukesGerakopoulos:2012pq}. Half of these surviving periodic orbits are stable and the other half are unstable.

The Poincar\'e section of a non-integrable Hamiltonian system has morphological features that are not present in an integrable system. The surviving stable periodic orbits appear as stable points on the section. These stable points are surrounded by so-called \emph{islands of stability} that fill a portion of the phase space, corresponding to regular orbits. Between the stable periodic orbits are the unstable periodic orbits, appearing as unstable points on the section. The asymptotic manifolds emanating from the unstable periodic orbits lead to a region of chaotic orbits that appear in the section as chaotic layers \cite{Lukes-Gerakopoulos:2014tga}. This arrangement on the section is known as a Birkhoff chain; see e.g.~Fig. 2 in \cite{LukesGerakopoulos:2012pq} for an illustration. 

In a \emph{weakly} perturbed non-integrable system, the islands of stability and the chaotic layers may fill up only a small fraction of the section, which is otherwise dominated by KAM tori. This makes it easy to overlook the Birkhoff chains, and thus the signature of non-integrability, in an automated search \cite{Dubeibe:2018tnv}. Fortunately, the signature of non-integrability is somewhat easier to detect near the Lyapunov orbit (LO) at the threshold between plunging and non-plunging orbits \cite{LukesGerakopoulos:2012pq}.

\subsection{Results}

We solved Hamilton's equations (\ref{eq:hamiltons}) numerically using Mathematica's {\tt NDSolve} for a range of initial conditions and parameters $E$ and $L$. We observed apparently-regular motion for large swathes of bound orbits when the orbital region is separated from the plunge region by the curve of zero velocity. However, we also observed key signatures of non-integrability. These signatures were seen most clearly for orbits near the Lyapunov orbit.

\begin{figure}
\begin{center}
 \subfigure[Two orbits in the meridian plane]{\includegraphics[width=7.8cm]{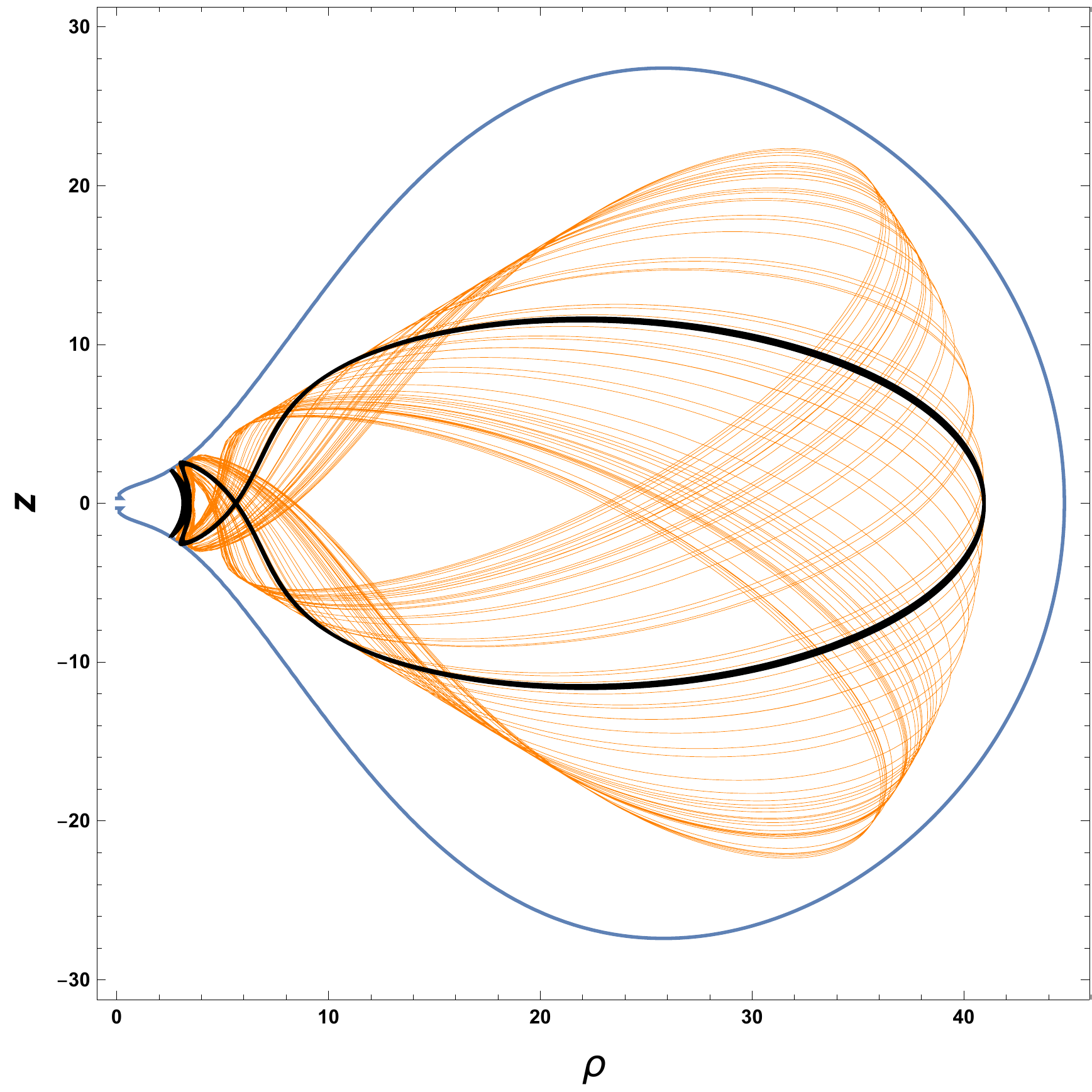}\label{fig:orbit}}
 \subfigure[Poincar\'e section $z = 0$]{\includegraphics[width=8.0cm]{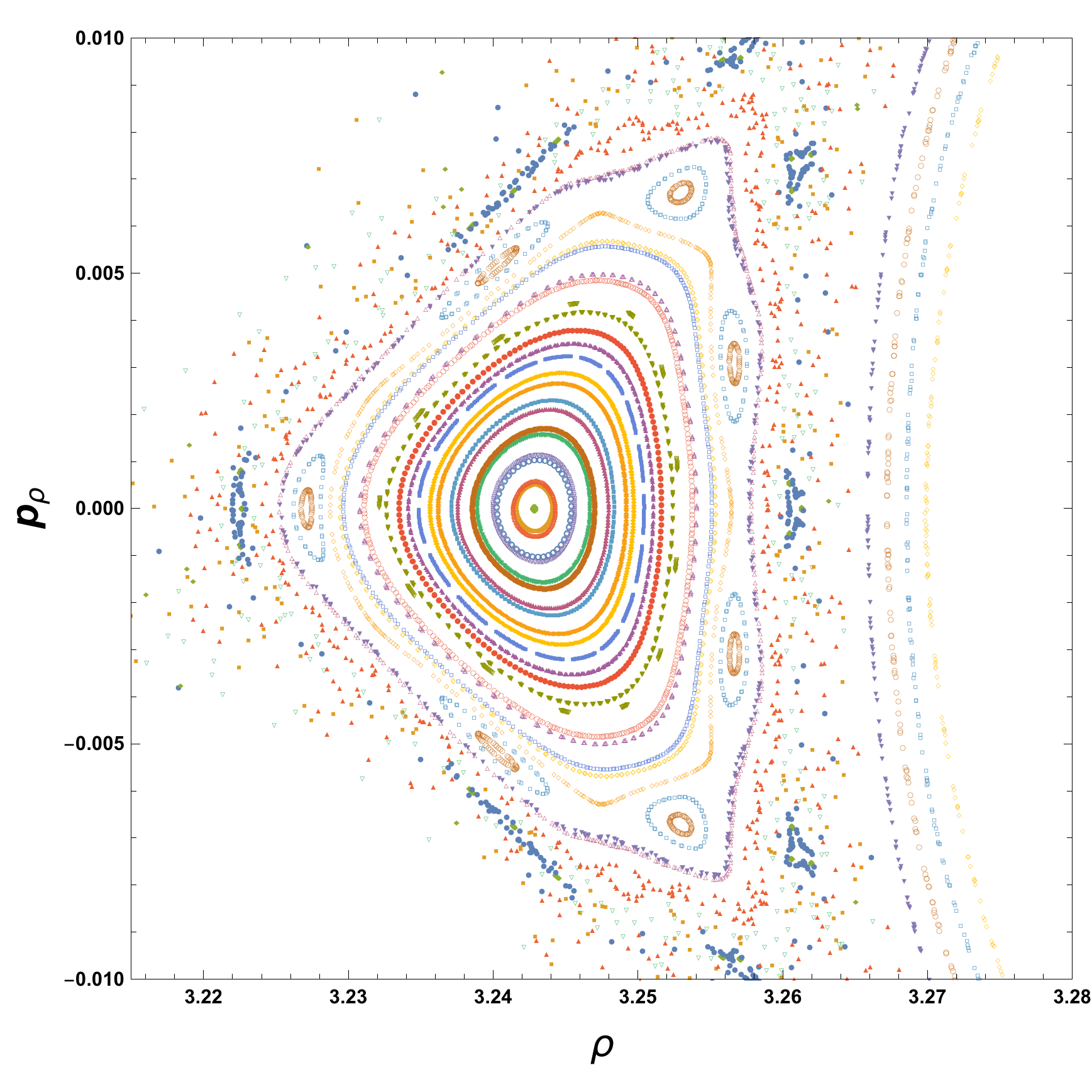}\label{fig:poincare}}
\end{center}
\caption{
Evidence of non-integrability in timelike geodesic orbits on the single-particle Chazy-Curzon spacetime, with $M=1$, $L = 3.0$ and $E = 0.98$. \emph{Left:} Orbits in the meridian plane $(\rho, z)$ with initial conditions $z=0$, $p_\rho=0$, $p_z > 0$, and $\rho = 3.2427$ (black, regular) and $\rho = 3.224$ (orange, irregular). The blue line is the curve of zero velocity. \emph{Right:} part of the $z=0$ Poincar\'e section, showing islands of stability and chaotic layers arranged into Birkhoff chains.
}
\label{fig:1}
\end{figure}

Figure \ref{fig:orbit} shows two example orbits in the meridian plane ($\rho, z$) for the choice $M = 1$, $E = 0.98$, $L = 3$. The black and orange orbits in Fig.~\ref{fig:orbit} start on the equatorial plane ($z=0$) with $p_\rho = 0$ at nearby positions, $\rho = 3.2427$ and $\rho = 3.224$, respectively. The black orbit is approximately periodic, and it generates a repeating sequence of points on the Poincar\'e section. Conversely, the orange orbit appears to be chaotic, and it generates a wandering sequence of points on the Poincar\'e section.

Figure \ref{fig:poincare} shows a part of the Poincar\'e section for the above choice of parameters, close to the Lyapunov orbit. We observe in Fig.~\ref{fig:poincare} the classic signatures of non-integrability, namely, islands of stability and chaotic layers arranged into Birkhoff chains. Chaotic orbits generate either (i) infinite sequences of points on the section, or (ii) terminating sequences once the orbiting body plunges towards $r=0$. 

\section{Conclusion}

The numerical results of Fig.~\ref{fig:1} are strong evidence that the single-particle CC spacetime cannot be Liouville integrable, and thus that there does not exist a fourth constant of motion. Further numerical evidence could be obtained by examining the `stickiness' of the rotation number, or the Fourier transform of the motion, or via other diagnostics of chaos. It is more challenging to obtain a definitive mathematical proof of non-integrability. One line of attack would be to extend the analysis of Maciejewski \emph{et al.}~\cite{Maciejewski:2013ncb}, which uses differential Galois theory to rule out the existence of any additional first integral meromorphic in the phase-space variables.

\section*{Acknowledgements}
With thanks to Jake Shipley for helpful discussions. I acknowledge financial support from the European Union's Horizon 2020 research and innovation programme under the H2020-MSCA-RISE-2017 Grant No.~FunFiCO-777740, and from the Science and Technology Facilities Council (STFC) under Grant No.~ST/P000800/1.

\bibliographystyle{apsrev4-1}
\bibliography{curzon.bib}

\end{document}